\RequirePackage{fix-cm}
\documentclass{svjour3}                     
\smartqed  
\usepackage{graphicx,color}
\usepackage{multirow}
\usepackage{amssymb}
\usepackage{amsmath}
\usepackage{xcolor}
\usepackage{afterpage}
\usepackage{hyperref}
\usepackage{float}
\usepackage{ulem,cite}
\begin{document}
\title{
Dynamical vortex production and quantum turbulence in perturbed Bose-Einstein condensates}

\titlerunning{Dynamical vortex production and quantum turbulence in perturbed BEC}        

\author{Lauro Tomio \and A. N. da Silva \and S. Sabari \and R. Kishor Kumar
}


\institute{
           Lauro Tomio \at
            Centro Internacional de Física,  I.F.,
              Universidade de Brasília, 70910-900 Brasília, DF, Brazil\\
              Instituto de F\'{i}sica Te\'orica, Universidade Estadual Paulista, 01140-070 S\~{a}o Paulo, SP, Brazil\\
                            \email{lauro.tomio@unesp.br}           
              \and            
              A. N. Silva 
              \at Instituto de F\'{i}sica Te\'orica, Universidade Estadual Paulista, 01140-070 S\~{a}o Paulo, SP, Brazil
               \\
              \email{anace.nunes@unesp.br}           
           \and
                         S. Sabari 
                         \at
              Instituto de F\'{i}sica Te\'orica, Universidade Estadual Paulista, 01140-070 S\~{a}o Paulo, SP, Brazil\\
                \email{sabari.subramaniyan@unesp.br} 
        \and  R. Kishor Kumar \at 
                      Instituto de F\'{i}sica Te\'orica, Universidade Estadual Paulista, 01140-070 S\~{a}o Paulo, SP, Brazil\\
                \email{kishor.bec@gmail.com} 
}

\date{Received: date / Accepted: date}

\maketitle

\begin{abstract}
Dynamical vortex production and quantum turbulence emerging in periodic perturbed 
quasi-two-dimensional (q2D) Bose-Einstein condensates are reported by considering 
two distinct time-dependent approaches. In both cases, dynamical simulations were
performed by solving the corresponding 2D mean-field Gross-Pitaevskii formalism.
(i) In the first model, a binary mass-imbalanced system is slightly perturbed by a
stirring time-dependent elliptic external potential. 
(ii) In the second model, for single dipolar species confined in q2D geometry, a 
circularly moving external Gaussian-shaped obstacle is applied in the condensate, 
at a fixed radial position and constant rotational speed, enough for the production
of vortex-antivortex pairs.  
Within the first case, vortex patterns are crystalized
after enough longer period, whereas in the second case, the vortex pairs remains 
interacting dynamically inside the fluid. In both cases, the characteristic 
Kolmogorov spectral scaling law for turbulence can be observed at some short time 
interval.
\end{abstract}

\section{Introduction}\label{Introduction}
{Within the going on experimental and theoretical Bose-Einstein condensates (BECs)
studies, in more recent years a lot of attention has been done on the possibilities to consider 
atomic BECs to get a more deep understanding on the classical well-known phenomenon known 
as turbulence~\cite{Kolmogorov}, in view of some similarities in the dynamics of vortices being 
generated in BECs~\cite{Barenghi2001,2002Vinen,2013Skrbek,2014Mantia}. Particularly, the interest 
was enhanced by the reported experiments in Ref.~\cite{Henn2009} on the observation of 
quantum turbulent regime in BEC. This interest also follows previous long-time 
investigations related to the Helium superfluidity phase transition, with the 
corresponding possible similarities with the Bose-Einstein condensate phase transition. 
In order to trace the related bibliography, several works and reviews are already available; among
the reviews, we can select Refs.~\cite{Tsatsos2016,2017Parker,Martin2017,Madeira2020b} as appropriate.\\
\\
The present report is contemplating two approaches in the dynamical vortex production in
BECs, which can lead to the so-called  {\it quantum turbulence}, by examining the 
corresponding incompressible part of the kinetic energy spectrum. Both approaches to 
produce vorticity in the system rely on the solution of extended Gross-Pitaevskii
formalisms, in which external periodic time-dependent interactions are included.
In the first approach, the dynamical production of stable vortices and quantum turbulence
is verified by assuming two component mass-imbalanced atomic condensates confined by a 
quasi-two-dimensional (quasi-2D) pancake-like trap potential slightly perturbed 
elliptically by a time-dependent periodic potential, with periodicity given by a parameter 
$\nu_E$~\cite{2023Silva}, which induces rotation in the confined system with final vortex
patterns being crystalized. This approach follows closely a previous 
study~\cite{Parker2005} in which a single condensate was considered; extended
to binary system as to verify possible effects due to mass-imbalanced atomic mixtures.
In the second approach we consider a condensed dipolar system, also confined in quasi-2D 
geometry, under the effect of an external circularly moving Gaussian-type penetrable 
obstacle, which is simulating a blue-detuned laser stirring mechanism. In this case, the
first task is to verify the critical frequencies of the obstacle in order to start
producing vortex pairs. By considering dipolar atoms, in a quasi-2D pancake-like confined 
condensate, another related purpose is to explore the possibilities to manipulate the
strength of the atom-atom interactions, from repulsive to attractive, via magnetic applied 
field, by considering the dipole orientations. As to manipulate the two-body interaction,
this approach came as an additional useful mechanism that can be used for dipolar 
systems, in addition to the already well-known procedure via Feshbach resonance techniques.
In both cases, the time evolution of the incompressible kinetic energy contributions is 
studied, within a semi-classical analysis, as to distinguish turbulent from non-turbulent
flows, guided by the corresponding expected classical behavior. Following these analyses,
transient turbulent regimes with the characteristic $k^{-5/3}$ Kolmogorov behavior are
identified in both model approaches, with the main difference being related to the 
transient time intervals at which turbulence behavior can be identified.
 
\section{Vortex patterns induced by periodic elliptical perturbation}\label{sec:2}
The vortex nucleation dynamics with corresponding characterization of turbulent flows 
ws recently studied in Ref.~\cite{2023Silva}, following previously related
studies~\cite{Parker2005}, by assuming a coupled mass-imbalanced BEC confined in quasi-2D
geometry perturbed elliptically.  For that, two easily accessible 
and controllable systems in cold-atom experiments were assumed, by considering mixtures of 
$^{85}$Rb-$^{133}$Cs and $^{85}$Rb-$^{87}$Rb, both in the miscible regime (implying that
the two species atom-atom scattering lengths $a_{ij}$, assumed being repulsive, are such
that $a_{12}^2<|a_{11}| |a_{22}|$). \\
\\ \noindent Next, we follow the essential dimensionless formalism provided in Ref.~\cite{2023Silva}, 
in which the units are given in terms of transversal trap frequency of the first species,
$\omega_\rho\equiv\omega_{1,\perp}$, with 
energies being in units of $\hbar\omega_\rho$, time in units of $1/\omega_\rho$, and
length units given by $\ell_\rho = 1\mu$m$ \approx 1.89\times 10^4 a_0$.
Within the quasi-2D pancake-like system, the 2D harmonic trap $V_0(x,y)\equiv (x^2+y^2)/2$,
is slightly perturbed elliptically by the time-dependent stirring
interaction: 
$V_s(x,y,t)=(\epsilon/{2})\big[(x^2-y^2) \cos(2\nu_E t)-2xy\sin(2\nu_E t)\big]$,
where $\nu_E$ is the laser stirring oscillating parameter, with $\epsilon$ the
corresponding strength. As one may find convenient, the total perturbed trap potential 
can be written in the usual polar coordinates, as
{\small \begin{eqnarray}
V_{\epsilon}(\rho,\theta,t;\nu_E)&=&
\frac{\rho^2}{2}\left[1+\epsilon\;\cos\left(2\theta+2\nu_E t\right)\right],
\label{2Dtrap}
\end{eqnarray}
}where $x=\rho\cos\theta$, $y=\rho\sin\theta$ and $\rho^2=x^2+y^2$.
As detailed in \cite{2023Silva}, the same formal expression for the confinement can be applied for both
species 1 and 2, by first adjusting their confining trap frequencies with the corresponding masses, 
such that ${m_2\omega_{2,\perp}^2}={m_1\omega_{1,\perp}^2}$. Therefore, conveniently, the
mass relations appear in the coupled formalism only in the kinetic energy term and the non-linear
interactions. The corresponding 2D coupled Gross-Pitaevskii formalism can be
expressed by
{\small \begin{equation}
\mathrm{i}\frac{\partial \psi_{i}  }{\partial t }=\Big[\frac{-m_{1}}{2m_{i}}{\nabla^2
}+V_{\epsilon}(\rho,\theta,t;\nu_E)+\sum_{j}g_{ij}|\psi_{j} |^{2}
\Big]\psi_{i},
\label{2d-2c}
\end{equation}
}where $-{\rm i}\nabla$ is the 2D momentum operator, with the two-component wave functions, 
$\psi_{i}\equiv \psi_{i}(\rho,\theta,t)$ being normalized to one, such that $\int d^2\rho\,|\psi _{i}|^{2}=1$.  
The parameters $g_{ij}$ are the 2D reduction of the corresponding three-dimensional parameters related to  
the intra- and inter-species two-body scattering lengths (respectively $a_{ii}$ and $a_{12}$), 
given by 
$g_{ij}\equiv \sqrt{2\pi\lambda}
\frac{m_1 a_{ij} N_j}{\mu_{ij}\ell_\rho},$
where $\mu_{ij} \equiv m_im_j/(m_i+m_j)$ is the reduced mass, $\lambda\gg 1$ is the shape-aspect parameter of the
pancake-like trap (assumed identical for both species), with $N_j$ being the number of atoms of
the species $j$.\\ \\
\noindent The perturbation is applied in the coupled condensate, as given in \eqref{2Dtrap}, by considering 
a very small strength $\epsilon\approx 0.025$, with oscillating parameter $\nu_E=1.25$, analogously
as in a previous \cite{Parker2005} with single component condensate, such that the original trap remains 
approximately with spherical format, besides the small time-dependent elliptical oscillation.\\ \\
\noindent In the dynamics, the system goes through different stages of instability, starting with 
shape deformation of the cloud, followed by symmetry breaking involving vortex nucleation,
with a final stage approaching rotating frame equilibrium, in which crystallization of
vortex patterns are verified. In order to characterize vortex nucleation in the coupled 
BEC system, a full analysis of vortex formation is performed in Ref.~\cite{2023Silva}, 
where the time evolutions of the relevant dynamical observables are presented for each
one of the two components of the mixture, including the total and kinetic energies, the
current densities, and torques. The kinetic energy injected is primarily compressible,
when the system is in the shape deformation regime,
switching to incompressible energy when vortex nucleation begins. 
The effective time-dependent rotational frequency of each component $i$, $\Omega_i(t)$, 
is estimated from the classical rotational relation, obtained from the expected values 
of the angular momenta and moment of inertia operators, as
\begin{eqnarray}
\Omega_i(t)&\equiv& \frac{\langle L_z(t)\rangle_i }{\langle I(t)\rangle_i }
=\frac{\int d^2\rho \;\psi_i^\star
\left( {-\rm i} \frac{\partial}{\partial\theta}\right)\psi_i} {\int d^2\rho 
|\psi_i|^2\;\rho^2}.
\label{rot-cl}\end{eqnarray} 
As shown in \cite{2023Silva}, by comparing two different mass-imbalanced mixtures, 
larger mass-imbalanced systems provide larger rotational frequency than lower 
mass-imbalanced mixtures, a result being reflected in the visible number of vortices 
shown in the final vortex patterns.
\begin{figure}[!ht]
\includegraphics[width=0.99\textwidth]{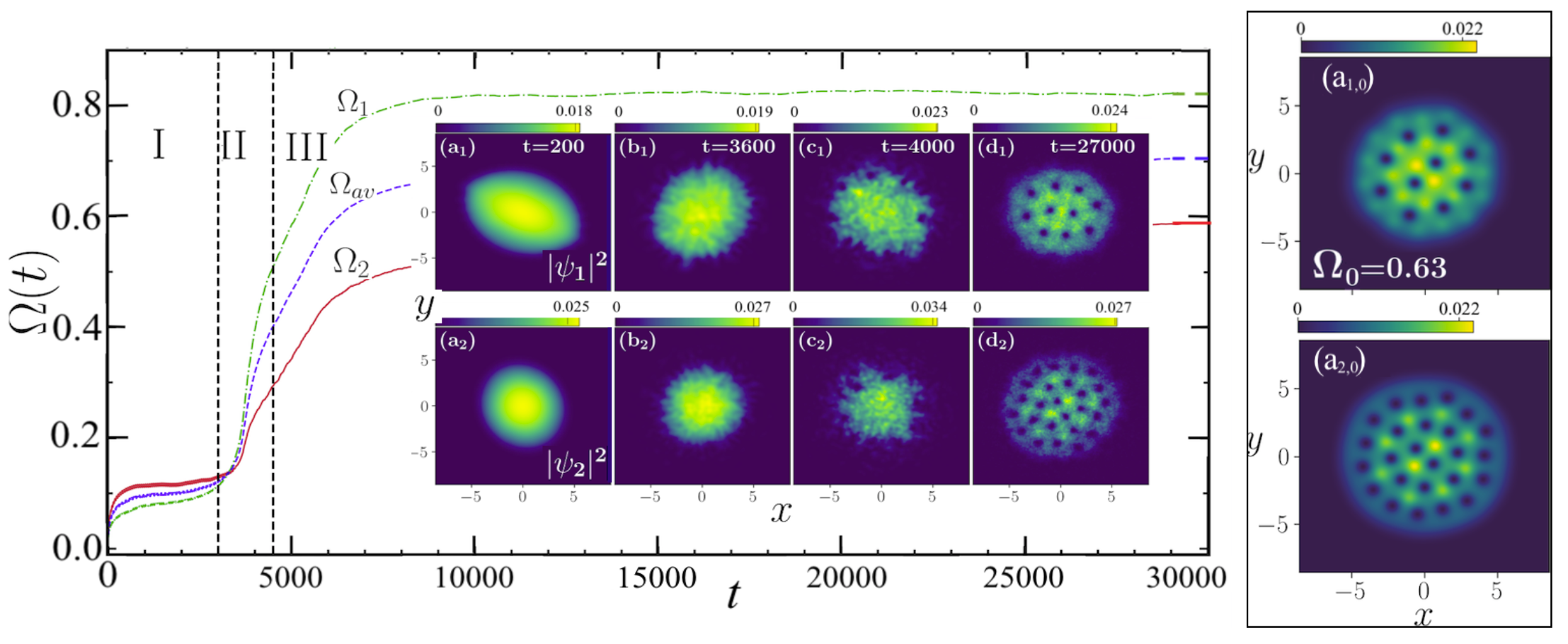}
\caption{(Color online)  The main panel shows the time evolutions of the rotational frequencies
$\Omega_1(t)$, for $^{85}$Rb (solid-red line); $\Omega_2(t)$,  for $^{133}$Cs (dot-dashed-green line); 
with the corresponding averaging, 
$\Omega_{av}(t)\equiv[\Omega_1(t)+\Omega_2(t)]/2$ (dashed-blue line). 
The vertical lines are approximately separating three time regimes: 
(I) shape deformation; (II) turbulent regime, with starting vortex nucleations; and 
(III) when vortex patterns  are settled.
In the inset of main panel, we also show time snap-shots for the two coupled
densities, $|\psi_1|^2$ [from (a$_1$) to (d$_1$)]
and $|\psi_2|^2$ [from (a$_2$) to (d$_2$)], corresponding to the three stages in the time
evolution. The intermediate turbulent region is represented by the panels (b$_{1,2}$) and
(c$_{1,2}$), respectively for $t=3600$ and 4000. The stirring parameters are $\epsilon=0.025$ 
with $\nu_E=1.25$.
In a separate box, in the right-hand-side, for a comparison with the final patterns 
(d$_i$), we have independent respective results (a$_{i,0}$) obtained from direct ground-state 
calculations of the GP equation, with the stirring interaction replaced by 
$ \Omega_0 L_z = -{\rm i}\Omega_0 \frac{\partial}{\partial \theta}$, assuming the same
$\Omega_0=0.63$ for both species. This figure is combining partial results presented in 
Ref.~\cite{2023Silva}.
All quantities are dimensionless, considering the units as defined in the text.
}
\label{fig01}
\end{figure}

\noindent In Fig.~\ref{fig01}, the main results for the dynamics are shown by considering the 
larger mass-imbalanced mixture studied in Ref.~\cite{2023Silva}, being $^{85}$Rb and 
$^{133}$Cs. In the principal panel, we present the long-time evolution results obtained 
for the time-dependent rotation $\Omega_i(t)$, together with some representative
snap-shot panels of the densities in the inset. The densities in the inset are
contemplating the three different
regimes of the evolution, with (a$_i$) being for the shape deformation stage, 
(b$_i$) and (c$_i$)  for the turbulent regime, with (d$_i$) for the final stage, with 
vortex-pattern crystallization. 
For comparison of the final vortex patterns obtained with the stirring perturbation, in a 
separate box of Fig.~\ref{fig01} we are also presenting results for the densities of the 
two species, obtained by direct  ground-state calculations of the GP equation, when the 
stirring interaction is replaced by 
$ \Omega_0 L_z = -{\rm i}\Omega_0 \frac{\partial}{\partial \theta}$. In this case, for
simplicity, we are assuming the same frequency $\Omega_0$ for both species, which 
is consistent with an averaging of the asymptotic limiting obtained in $\Omega_i(t)$.
\begin{figure}[!h]
\includegraphics[width=11.cm,height=4.5cm]{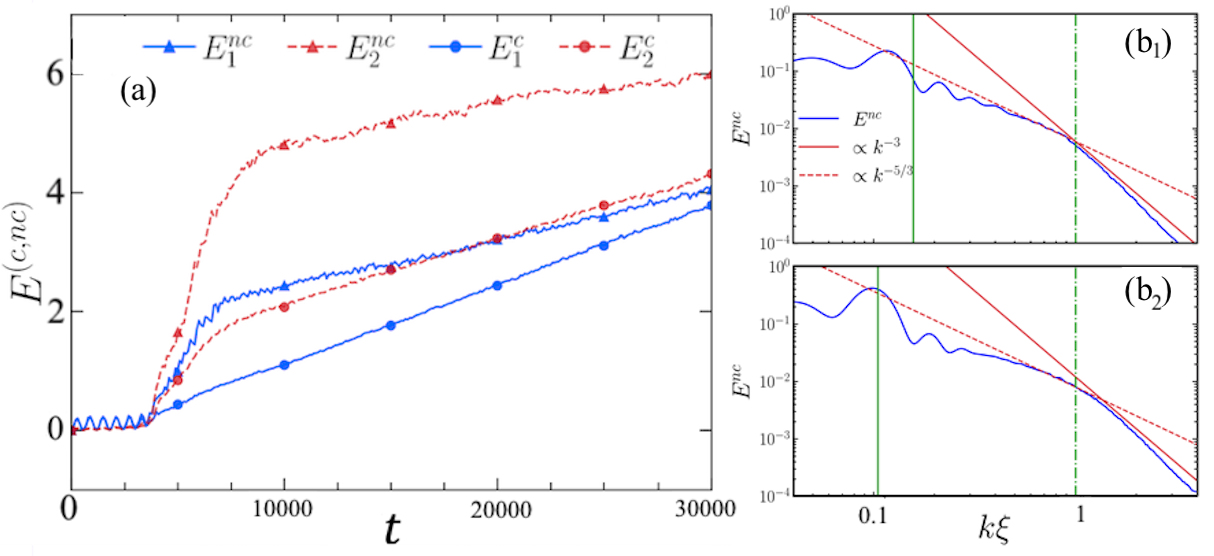}
\caption{(Color online)  
In panel (a),for each component of the mixture, $^{85}$Rb ($i=$1, solid-blue lines) and $^{133}$Cs 
($i=$2, dashed-red lines), the time evolutions are presented (in the laboratory frame) for 
the compressible and incompressible parts of the kinetic energies, respectively, $E^{c}_i$ 
(indicated with bullets) and $E^{nc}_i$  (indicated with triangles).
In panels (b$_1$) and (b$_2$) we show, respectively, 
the incompressible kinetic energy spectra, $E^{nc}\equiv E^{nc}(k,t)$, for 
the $^{85}$Rb-$^{133}$Cs mixture, obtained by averaging over 50 samples in the turbulent 
time-interval (II) indicated in Fig.~\ref{fig01}, which approximately agree with the classical 
Kolmogorov $k^{-5/3}$ power-law behavior in the shown interval $0.1<k\xi<1$ (red-dashed lines), 
being modified to the $k^{-3}$ in the ultraviolet region. 
In the stirring perturbation, the parameters are $\epsilon=0.025$ with $\nu_E=1.25$.
This figure is combining partial results reported in Ref.~\cite{2023Silva}.
All quantities are dimensionless, considering the units as defined in the text.
}
\label{fig02}
\end{figure}

\noindent By verifying possible characterization of turbulent behavior in the dynamics of the 
coupled condensate, the results obtained for the incompressible kinetic energy spectra 
of the two species were investigated, with the corresponding main results resumed in 
the three panels of Fig.~\ref{fig02}. When computing the incompressible kinetic energy, 
as a measure for the vortex energy, a care was taken to subtract the uninteresting rigid 
body velocity field associated with the rotation. The time evolutions for the 
compressible (c) and incompressible (nc) parts of the kinetic energies are shown in the 
main panel, in which the starting time of the turbulent regime (near $t\approx 3000$) 
can be identified by the splitting between the two parts of the kinetic energies for 
both species. See also by the vertical dashed line between regions I and II, in the main 
panel of Fig.~\ref{fig01}. 
As noticed by the results shown in the panels (b$_1$) and (b$_2$) of Fig.~\ref{fig02}, 
in this regime of the time evolution, the incompressible kinetic energy presents  
evidence of the Kolmogorov $k^{-5/3}$ scaling, for values of the $k\xi$ momenta (where 
$\xi$ is the healing length corresponding to the respective species $i$) 
smaller than 1, which is modified to a $k^{-3}$ power-law behavior in the ultraviolet 
regime determined by the vortex core structure~\cite{2012Bradley}. 
While the Kolmogorov scale range appears somewhat limited, the spectrum is 
consistent with energy transport to large scales as disordered vortices enter the 
system and begin to organize. 
The vortices in this stage are close to the boundary of the condensate, and the 
Kolmogorov power-law scaling vanishes once stable vortex configurations develop. 
As comparing two kind of mass-imbalanced systems in \cite{2023Silva}, for
$^{133}$Cs$^{85}$Rb and $^{85}$Rb$^{87}$Rb, it was noticeable that larger mass-imbalance 
system shows the approximate $k^{-5/3}$ power-law behavior for a longer 
time window. Also, the dynamical production of stable patterns of vortices is verified 
to be a much faster process for larger mass-imbalanced systems than for smaller 
mass-imbalanced ones.\\ \\
\noindent 
In the following section, instead of a small time-dependent perturbation, another 
approach is reported to generate vorticity in a dipolar BEC system, by considering 
direct external obstacle moving periodically inside a single condensed system. 

\section{Vortex-antivortex dynamical production in dipolar BEC under circularly moving obstacle}\label{sec:3}
In this section, within another approach which is being considered experimentally to produce vorticity in a 
condensate~\cite{Lim2022}, an external penetrable Gaussian-type moving obstacle is considered to generate
 vortex-antivortex pairs, following a recent study detailed in \cite{2024Sabari}, in which Bose-Einstein 
 condensed dipolar atoms were considered.  
 Applied to a dipolar condensate confined by a quasi-2D pancake-like harmonic trap, 
 two variants of a circularly  moving Gaussian-shaped penetrable obstacle were assumed in~\cite{2024Sabari}, 
 following previous related investigations with linearly-moving obstacle~\cite{2017Sabari,2018Sabari}. 
 In the first variant, the obstacle moves with constant
rotational frequency at a fixed given radius inside the condensed fluid, with its amplitude 
$A_0$ assumed to be close to 90\% of the stationary chemical potential $\mu$. 
For the second variant, it was verified the effect of an additional dynamics provided by a 
vibrating amplitude with frequency larger than the stirring rotational one.
By having in mind that the most significant results can be appreciated through the simpler 
model, here we discuss the case in which the shape parameters (amplitude and width) of the circularly 
moving obstacle remain constant in the condensate. Within this approach, vortex-antivortex pairs
are generated by the obstacle, once the rotational speed is larger than a critical threshold one.
Therefore,  instead of the stirring time-dependent perturbation, here we assume a penetrable 
Gaussian-shaped obstacle, moving circularly at a fixed radius $r_0$ and rotation speed $\nu$, having 
amplitude $A_0$ and width $\sigma$, 
{\small \begin{eqnarray}\label{VG}
V_{G}(x,y,t)& \equiv &A_0 \exp\left(-\frac{\left[x-x_0(t)\right]^2+
\left[y-y_0(t)\right]^2}{2\sigma^2}\right),
\end{eqnarray}
}where $x_0(t)\equiv r_0 \cos(\nu \,t)$ and $y_0(t) \equiv r_0 \sin(\nu \,t)$ are the instant positions 
of the obstacle in the 2D plane, with $\sigma$ being the corresponding standard radial deviation
(close to half-width of the distribution). \\ \\
\noindent
The study of vorticity in BEC was extended from non-dipolar to dipolar atomic systems motivated by several
previous studies followed by different experimental realizations of dipolar BECs in recent years~\cite{Griesmaier2006,Lahaye2007,2008Koch,Lu2011,Aikawa2012,2022Klaus,2022Miyazawa},
in which the atoms have large magnetic dipole moments, as discussed in the topical review~\cite{Martin2017}. 
The interest, in this case, relies on the fact that dipolar condensates can combine properties of superfluidity 
(having quantized circulation vortices) with ferrofluidity, in which the long-range dipolar
interactions can induce new effects due to the interplay of magnetism with vorticity.
In response to the magnetic
field, basic properties of vortex and vortex patterns are expected to be affected, leading to shape deformations
and/or new patterns.
Dipolar condensates are also opening new possibilities to  manipulate the atom-atom interactions,
not only through Feshbach resonance mechanism, but also by controlling of magnetic moment orientations 
between atoms in relation to the trap geometry. 
Along the present communication, before submitting the system to a circularly  moving obstacle, 
we assume net repulsive contact plus dipolar nonlinear interactions, in order to keep the system in stable 
configuration. 
By following~\cite{2012Wilson}, for the formalism reduction from three- to 2D, as also
detailed in~\cite{2017KumarJPC,Kumar2017,Kumar2019,2024Sabari}, with  
the dipolar term in the Fourier-transformed momentum space,  the effective equation for the dipolar BEC 
is given by
{\small \begin{eqnarray}
 {\rm i}\frac{\partial \psi}{\partial t}&=&\left\{
 -\frac{1}{2}\nabla_{\rho}^2+\frac{\rho^2}{2}+ V_{G} +g_{s}|\psi|^2 
 +g_{dd}\int \frac{d^2k_{\rho}}{4\pi^2}
e^{i \mathbf{k}_{\rho}.\tilde{\rho}}\tilde{n}(\mathbf k_{\rho})
\widetilde{{V}}^{(d)}({\bf k_\rho}) \right\} \psi,
\label{gpe_scaled}
\end{eqnarray}
}which corresponds to the single-condensate formalism version of \eqref{2d-2c} for dipolar system, with
$\psi\equiv\psi(\boldsymbol{\rho}, t)$ normalized to one, $g_s\equiv \sqrt{8\pi\lambda} {a_s N}/{\ell_\rho}$
($N$ the number of atoms) and $V_G\equiv V_{G}(\boldsymbol{\rho},t)$ given by \eqref{VG}.
In the above 2D Eq.~\eqref{gpe_scaled}, $\tilde{n}(\mathbf k_{\rho})$ and $\widetilde{{V}}^{(d)}({\bf k_\rho})$
are the Fourier transforms of the 2D density and dipolar potential, respectively. As shown in Refs.~\cite{2024Sabari},
as well as by following Refs.~\cite{2012Wilson,Zhang2016}, the dipole-dipole interaction in momentum space, 
after an averaging of the polarization rotating field in the $(k_x,k_y)$ plane,  can be expressed by 
{\small \begin{equation}
{\widetilde{V}}^{(d)}({\bf k_\rho}) = \frac{3\cos^2\alpha-1}{2}
\left[2-3\sqrt{\frac{\pi}{2\lambda}}k_\rho
\exp{\left(\frac{k_{\rho }^{2}}{2\lambda}\right)} 
{\rm erfc}\left( \frac{k_{\rho }}{\sqrt{2\lambda}}\right)\right] 
\label{DDI-2D}
\end{equation} 
}where ${\rm erfc}(x)$ is the complementary error function of $x$,
with $\alpha$ defining the inclination of the dipole moments relative to the
perpendicular $z-$direction.
Therefore, as show above, the strength of the atom-atom long-ranged 
dipolar interaction is guided by the factor $g_{dd}\left(3\cos^2\alpha-1\right)/2$
(which goes from $g_{dd}$ for $\alpha=0$ to $-\frac{g_{dd}}{2}$ for $\alpha=90^\circ$,
being cancelled out for $\alpha\approx54.7^\circ$. )
\\ \\
\noindent
In the dynamical vortex production simulation, studied in \cite{2024Sabari},
the circularly moving obstacle inside the condensed fluid was assumed with frequency 
$\nu$, at a given fixed radius $r_0$ (which, obviously is restricted to the size
of the condensate), such that the obstacle velocity is $v=\nu r_0$. 
Given that, by first observing that
the corresponding critical velocity $v_c=\nu_c r_0$ to produce vortex-antivortex
pairs remains close to a constant value when varying $\nu$ and $r_0$,
one can verify that $r_0$ cannot be near the maximum unperturbed trapped density ($r_0=0$),
where vortex production needs unrealistic too-large values of $\nu$, as well as not 
close to the low-density region of the condensate. 
By assuming an intermediate value, $r_0=3.5\ell_\rho$, the critical
rotation  was verified being near $\nu_c=0.6\omega_\rho$, when considering the
Gaussian obstacle parameters $A_0=36\hbar\omega_\rho$ and $\sigma=1.5\ell_\rho$,
for the nonlinear short-ranged contact interactions (SCI) fixed by the
scattering length $a_s=50a_0$, with the corresponding dipole-dipole interactions
(DDI) fixed by $a_{dd}=66a_0$, with $\alpha=0$. The choice of $a_s$ and $a_{dd}$
parameters were directly associated to the dipolar erbium isotope $^{168}Er$.
Therefore, the observed dynamical results reflect effects due to both nonlinear
atom-atom contact and dipolar interactions, which can be distinguished by 
switching off one of them. As known, the SCI can be modified by using Feshbach 
resonance techniques to increase or decrease $a_s$. Similarly, the DDI can
be modified by manipulating the polarization angle $\alpha$, being reduced to zero 
for $3\cos^2\alpha=1$ ($\alpha=\alpha_M\approx 54.7^\circ$, known as the magic angle),
as well as changing the DDI to be attractive for $54.7^\circ<\alpha< 90^\circ$.
\\ \\
\noindent
As a complement to the dynamical production of vortex-antivortex pairs investigated in 
Ref.~\cite{2024Sabari}, as well as to appreciate the effect of the long-ranged DDIs in 
contrast to some well-known effects verified with pure the short-ranged contact interactions,
we show in Fig.~\ref{fig03} a few set of sample results obtained in our simulations, 
related to the vortex production and their dynamics, by considering {\it pure-}contact (when DDI=0)
or {\it pure-}dipolar (when SCI=0) effects. For this comparison, taken the Gaussian obstacle with
identical parameters in both cases, we need to adjust the parameters such that in both the cases 
we have similar condensates with about the same root-mean-square (rms) radius and chemical potentials. \\
\noindent By taking as reference the dipolar condensate with $^{168}Er$, the parameters and physical
quantities considered in Fig.~\ref{fig03} are provided in the following Table~\ref{tab:1}.
\begin{table}[h]
\caption{Parameters and physical quantities assumed in the simulations shown in 
Fig.~\ref{fig03}.
The  Gaussian obstacle, 
which is circularly moving with frequency $\nu=0.8\omega_\rho$,
at a fixed radial position $r_0=3.5\ell_\rho$ in the condensate, have identical parameters
in all the panels, ($a_{j}$)  and ($b_{j}$),
represented in Fig.~\ref{fig03}, being $A_0=36 \hbar\omega_\rho$ and $\sigma= 1.5 
\ell_\rho$.}  \label{tab:1} 
\begin{tabular}{|c|c|c|c|c|c|c|c|c|c|}
 \hline\hline
Description & $a_s$ & $(a_{dd},\alpha)$ & $\mu (\hbar\omega_\rho)$ & $E (\hbar\omega_\rho)$
&$\sqrt{\langle \rho^2\rangle}(\ell_\rho)$&$ |\psi |^2_{max}(\ell_\rho^{-2})$ \\  
 \hline \hline
 SCI=0 ($a_{j}$) & 0            & $(91a_0,0^\circ)$  &47.39 & 31.49 & 5.58 & 0.007 \\  
 \hline 
 DDI=0 ($b_{j}$) & $177a_0$& $(66a_0,54.7^\circ)$&47.37 & 31.60& 5.62 & 0.007 \\
 \hline\hline 
 \end{tabular}
\end{table}
\begin{figure}[!ht]
\centering
\includegraphics[width=0.9\textwidth]{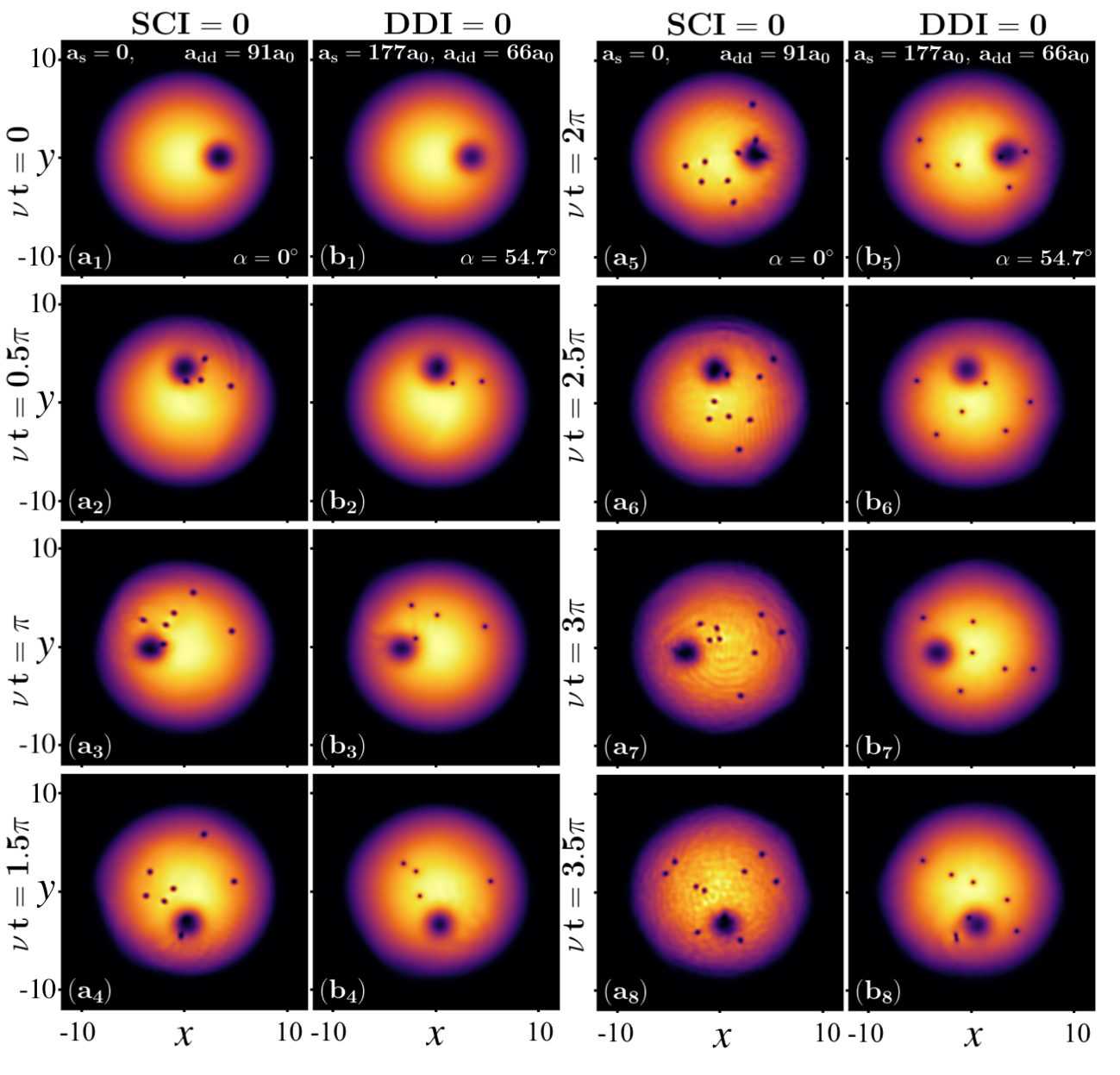}
\caption{Comparative snap-shot results of densities $|\psi(x,y,t)|^2$ 
(units defined in the text) obtained in 
the dynamics of vortex-antivortex production, for the circularly rotating 
obstacle (large shadowing circle) at $r_0=3.5\ell_\rho$, 
with frequency $\nu=0.8\omega_\rho$.
As indicated, the panels (a$_j$) and (b$_j$), 
in the left two columns, are from $\nu t=0\; (j=1)$ upto $1.5\pi\; (j=4)$; 
and, in the right two columns, from $\nu t=2\pi\;(j=5)$ upto
$3.5\pi\;(j=8)$.
In the first and third columns we have pure dipolar interactions (SCI=0), 
with $a_{dd}=66a_0$ ($\alpha=0$) and $a_s=0$. In the second and forth columns are for
pure contact interactions (DDI=0), with $\alpha=54.7^\circ$ and $a_s=177a_0$.
Density levels vary from zero (black) to maximum (yellow),
with the small black circles being the vortex and antivortex (produced in pairs), which are moving inside the fluid.
}
\label{fig03}
 \end{figure}

\noindent The results for zero nonlinear contact interactions (SCI=0) are represented in the eight 
panels ($a_{j}$) of Fig.~\ref{fig03}, implying results for {\it pure-dipolar} 
long-ranged interacting systems.  They are in the first and third columns of Fig.~\ref{fig03}, with $\nu t$ 
varying anti-clockwise from 0 to 3.5$\pi$,
as indicated. For the case of DDI=0, we simply adjust the orientation angle such that $\alpha=54.7^\circ$, 
increasing the repulsive contact interaction using $a_s=177a_0$, such that the chemical potential and
rms values remain fixed. Similarly, the results are shown in the eight panels (b$_{j}$), in 
the second and forth columns of Fig.~\ref{fig03}, with the same variation of $\nu t$. 
Therefore, the associated animated results can be visualize comparatively in this
figure, such that one can separate the short-ranged and long-ranged net effect interactions 
in the production and dynamical behavior of vortex-antivortex pairs inside
the fluid.  In this dynamics, differently from the case that the obstacle moves linearly passing through the 
center~\cite{2018Sabari}, the vortex and antivortex of a single pair emerge at slightly different times, 
which can be understood by the fact that the fluid in one side of the moving obstacle is denser 
than the opposite side. For an anti-clockwise circular movement of the obstacle, a clockwise vortex 
emerges (at the right side of the obstacle) at a slightly shorter time than the corresponding anti-clockwise 
anti-vortex. A related interesting aspect, observed in the dynamics of vortex-antivortex in the fluid, is their tendency 
to persist longer time moving inside the fluid without being dissipate or annihilate, in which 
the non-annihilation can be related to the fact that both vortex and anti-vortex of a given pair 
emerge with slightly different kinetic energy in a superfluid with almost no viscosity.\\ 
\\
\noindent It is being noticeable in this comparison of pure-DDI condensate with pure-SCI condensate
that the long-range repulsive dipolar interactions are affecting not only the subsequent 
dynamics of the vortex pairs interacting inside the fluid, but also the initial emission
of vortex pairs, indicating that the critical velocities in case of pure dipolar system are slightly 
lower than in the case of pure contact interactions. In principle, one could expect the initial
vortex production relying mainly in the obstacle position and its shape. But apparently this
is not the case, which can be attributed to the different fluid characteristics related to
range interactions. On this regard, as discussed in Ref.~\cite{Martin2017}, following cited
related works, by comparing vortex-free solutions with DDI, using Thomas-Fermi approximation
in the general 3D case, it was noted that, besides the fact that non-dipolar BECs have the same 
aspect ratio as given by the harmonic trap, for dipolar case the aspect ratio is modified.
In view of these observations, a more detailed analysis is required
by considering different increasing rotational velocities of the obstacle.\\   
\\ 
\noindent As verified in the dynamics represented by the snap-shots given in panels (b$_j$) of Fig.~\ref{fig03}, 
the condensed cloud remains approximately uniformly distributed in case with pure SCI,
without DDI, even after the started production of vortices. However, within a pure dipolar system (no SCI), 
shown in panels (a$_j$) of Fig.~\ref{fig03}, one can observe not only the slightly increasing number of 
vortex pairs, but also the production of stripes and fluctuations in the fluid, which can be taken as
clear signs of the long-range interactions between atoms.  More clearly, the effect of long-ranged DDI 
 as compared with the short-ranged SCI, can be appreciated in the second loop ($\nu t \ge 2\pi$) shown 
 in Fig.~\ref{fig03}.
These results are also compatible with experimental observations~\cite{2022Klaus}, as well as in line with 
previous numerical calculations in~\cite{2006Yi,2009Wilson}, performed 
for 2D pancake-like dipolar condensate where, for perpendicular polarized dipoles they observe, 
in case of a single vortex, that density ripples form about the vortex core for large trap ratios 
($\lambda\sim 100$). 
This kind of density behavior can also be observed within our results when more vortices are 
emerging inside the fluid.
 In case of pure DDI, it is noticeable the occurrence of stripes and vortex ripples, 
 which are clearly in contrast with the SCI case. \\
\\ 
\noindent These different rich dynamics, observed by comparing the short-ranged with long-ranged atom-atom
interactions in a condensate, as well as the subsequent interactions between vortices and antivortices
inside the fluid, deserve more detailed analyses by simulating different conditions and obstacle velocities.
In the present contribution, our study will be limited to the above discussion related to our direct 
results obtained from the solution of the GP formalism, and by the following analysis of
possible quantum turbulence behavior.

\begin{figure}[!ht]
\centering\includegraphics[width=1.0\textwidth]{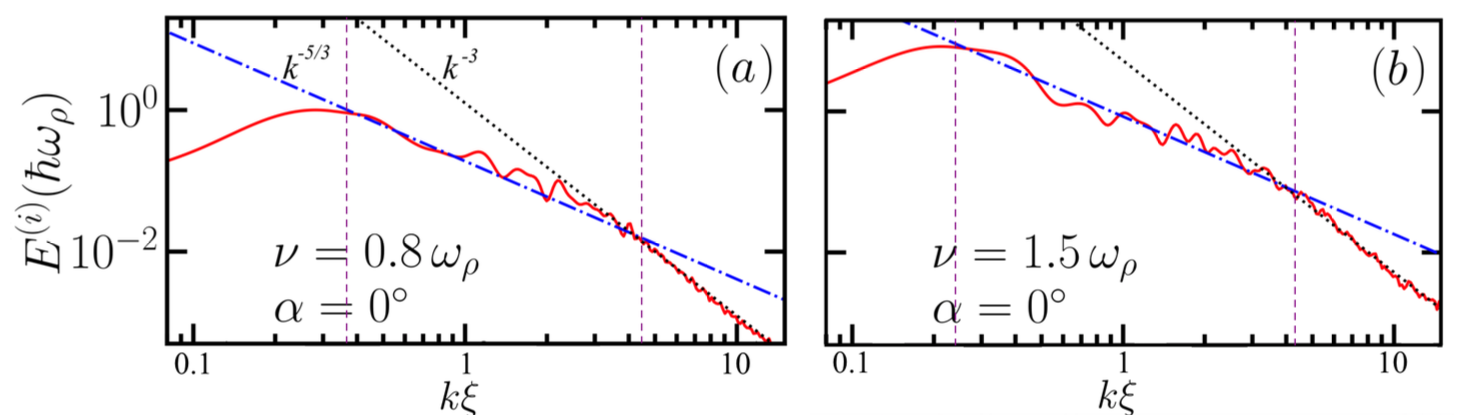}
\caption{Incompressible kinetic energy spectrum $E^{(i)}$ as function of the
the momentum $k$ (in units of the healing length $\xi$), for two rotational
frequencies of the obstacle [$\nu=0.8\omega_\rho$ (a) and $1.5\omega_\rho$ (b)], 
located at $r_0=3.5\ell_\rho$,
considering $\alpha=0^\circ$ (where repulsive DDI is maximized). 
The results are obtained by time averaging over 10 samples within the interval of vortices
emission ($2< \omega_\rho t< 10$). As shown, the Kolmogorov-kind behavior $k^{-5/3}$, indicated
between the vertical dashed lines, starts at lower values of $k$ for higher obstacle rotation.
The contact and dipolar interactions are set to $a_s=50a_0$ and $a_{dd}=66a_0$,
respectively.}
\label{fig04} 
 \end{figure}

\noindent By considering the kinetic energy spectrum, along the same procedure
as done in Ref.~\cite{2023Silva}, as outlined in the previous section 2,
a time interval is also verified in such case with the incompressible part of the 
kinetic energy behavior following approximately the Kolmogorov classical law. 
This dynamical process occurs in this case within a shorter time interval
in which the vortex pairs are being emitted. As the vortex pairs are not produced at 
the same time, in the present case, summarized in two panels of Fig.~\ref{fig04},
we consider two different velocities of the obstacle, given by $\nu=0.8\omega_\rho$
 [panel (a)] and $\nu=1.5\omega_\rho$ [panel (b)]. The behavior of $E^{(i)}$, as function
 of $k\xi$ (where $\xi$ is the healing length), is extracted from time averaging 
 the corresponding time-dependent behavior, in the time interval started when the 
 incompressible part of the kinetic energy  deviates from the corresponding compressible 
 part, within a short time interval till vortex pairs have been dynamically interacting
 in the fluid. As also verified in the present case, the classical Kolmogorov power-law 
 behavior $k^{-5/3}$ is being characterized for the two different velocities of the
 obstacle, providing indication for {\it quantum turbulence}, within the momentum spectral 
 analysis of the incompressible kinetic energy.  On this regard, our analyses are still 
under more detailed investigation, considering different combinations of dipolar 
and contact interactions, and also by assuming different rotational frequencies of the 
obstacle inside the fluid. As one can noticed by
a visual analysis of the dynamical production and vortex-antivortex interactions 
inside the fluid, as shown in Fig.~\ref{fig03} for the limited two cycles with
$\nu=0.8\omega_\rho$, the dipolar fluid starts to present stronger fluctuations and stripes 
affecting the vortex-antivortex dynamics, which may require further investigation. 
It should also be noticed the quite interesting aspects previously revealed in 
Ref.~\cite{2018Bland} of quantum turbulence arising in dipolar Bose gas
 from highly non-equilibrium thermal states, without external forces. In such 
 case, quantum turbulence emerges strongly polarized when the contact 
and dipolar interactions are of the same order.
As considering our present case, with an external circularly moving obstacle applied to
the Bose system, a more detailed comparative analysis of non-dipolar and dipolar systems 
can be implemented by varying the obstacle speed, in which for higher speeds vortex 
clusters can also be emitted.

\section{Conclusions}\label{sec:4}
Within a comparative analysis, two recent investigations are reported in this communication,
considering production of vorticity with eventual turbulence behavior. 
In section 2, a binary mass-imbalanced BEC system under periodic small elliptically perturbation 
was analyzed by considering some results extracted from Ref.~\cite{2023Silva} for a particular 
large mass-imbalanced coupled system.
In section 3, another approach was considered for vortex-antivortex production, with a single 
dipolar condensate under circularly periodic Gaussian-shaped obstacle, simulating a laser stirring 
mechanism.
In both the cases, the dynamics of vortex production is further explored by kinetic energy 
spectral analysis, in which the characterization of turbulent dynamics is verified 
in a time regime preceding the stable vortex formations. As follows from classical fluid dynamics, 
it is understood that vortex tangles are usually signatures of turbulence associated with the 
flow of incompressible viscous fluids.
Therefore, by concentrating the spectral analysis on the incompressible 
kinetic energy part, obtained by averaging over several samples in the time 
evolution, the characterization of the turbulence behavior was possible to be 
established by verifying that the incompressible kinetic energy $E^{(i)}(k)$ 
follows approximately the classical Kolmogorov power-law $k^{-5/3}$~\cite{Kolmogorov} 
in the momentum region $k\xi\lesssim 5$ (where $\xi$ is the healing length), 
changing to $k^{-3}$ as $k$ goes to the ultraviolet region, consistent with previous 
studies~\cite{2012Bradley}. 
A common feature in our analysis is that the time interval observed for the turbulence 
behavior is initiated when the compressible and incompressible parts of the kinetic energy 
start deviating from each other.
In the case of small elliptically perturbation discussed in section 2, this happens when vortex 
formations start migrating from low-density regions to the internal part of the condensate. 
However, for the case in which the condensate is being affected by circularly moving obstacle, 
as  analyzed in section 3, such dynamics occurs immediately before vortex-antivortex pairs 
are being produced. 
In both cases, such turbulent behavior occurs during a limited time interval as compared with
the full period of vortex dynamics, but with the Kolmogorov spectral law being verified within similar 
intervals in momentum space, as seen in panels (b$_i$) of Fig.~\ref{fig02} and Fig.~\ref{fig04}.  
In the case the fluid is under the influence of a moving obstacle, the dynamics can be modified
significantly by varying the velocity of the obstacle, also with production of vortex clusters, it is
also required a more detailed investigation in order to establish possible 
time intervals where Kolmogorov criterium may be applied.\\
\\ 
\noindent
Apart from the above discussed turbulent behavior, when assuming a pure-dipolar condensate 
under rotating Gaussian-shaped obstacle, it is also shown a rich dynamics that occurs
inside the condensate, which is quite different from the dynamics following the case
of pure-contact nonlinear interactions.
This dynamics, which follows the vortex-antivortex production, is exemplified in a specific
case in which the rotational frequency is $\nu=0.8\omega_\rho$ of the obstacle
at a fixed radial position $r_0=3.5\ell_\rho$. 
It is represented by a series of eight pair of panels, considering two extreme cases 
as related to the nonlinear interactions: One case, in which the contact interactions are set to zero
($a_s=0$, or SCI=0), such that the non-linear interactions are purely due to dipolar interactions;
with the other case in which dipolar interactions are cancelled out, by taking 
$\alpha=54.7^\circ$ (DDI=0). In this case, the SCI is given by $a_s=177a_0$, in order to
keep both condensates having the same rms radius and chemical potentials.
Within these comparative simulations, the vortex-antivortex dynamics became clearly 
evidenced in both the cases (pure dipolar, or pure contact interactions). By considering
the dynamical movement and interactions of the vortices and antivortices inside the fluid,
as well as the observed density fluctuations, one can clearly verify the richer dynamics
of dipolar systems in comparison with the case that dipolar interactions are cancelled out. 
Finally, relevant to say is that all the interesting observed dynamics inside the fluid
are verified within the simpler mean-field approach with cubic non-linear interactions,
without the addition of other nonlinearities brought by quantum fluctuations.
\begin{acknowledgements}
We acknowledge partial support from Fundação de Amparo à Pesquisa do Estado de 
São Paulo [Contracts No. 2020/02185-1 (S.S.) and No. 2017/05660-0 (L.T.)], 
Conselho Nacional de Desenvolvimento Científico e Tecnol\'ogico  
(Procs. 304469-2019-0  and 464898/2014-5) (L.T.), 
Coordena\c c\~ao de Aperfei\c coamento de Pessoal de N\'\i vel Superior (A.N.S.),
and  Marsden Fund (Contract No. UOO1726) (R.K.K.). 
\end{acknowledgements}


\end{document}